\documentclass[12pt,onecolumn,conference]{IEEEtran}
\usepackage{epsf,times,amsmath,color}
\usepackage{graphicx}
\usepackage{amsfonts}
\usepackage{amsbsy}
\usepackage{amsmath}
\usepackage{acronym}
\usepackage[IEEE]{mymaths}
\usepackage{tikz}
\usepackage{cite}
\usepackage{multirow}
\usepackage{subfigure}

\renewcommand{\B}[1]{\mathbf{#1}}

\acrodef{AWGN}{additive white Gaussian noise}
\acrodef{BER}{bit error rate}
\acrodef{BPSK}{binary phase shift keying}
\acrodef{CDF}{cumulative distribution function}
\acrodef{CP}{cyclic prefix}
\acrodef{CSI}{channel state information}
\acrodef{DMT}{discrete multitone}
\acrodef{FER}{frame error rate}
\acrodef{FFT}{fast Fourier transform}
\acrodef{GSVD}{generalized singular value decomposition}
\acrodef{ICI}{interchannel interference}
\acrodef{IFFT}{inverse fast Fourier transform}
\acrodef{ISI}{intersymbol interference}
\acrodef{KKT}{Karush-Kuhn-Tucker}
\acrodef{LDPC}{low-density parity check}
\acrodef{LOS}{line of sight}
\acrodef{MAP}{maximum a posteriori}
\acrodef{MIMO}{multiple input multiple output}
\acrodef{MMSE}{minimum mean square error}
\acrodef{MSE}{mean squared error}
\acrodef{NLOS}{non line of sight}
\acrodef{OFDM}{orthogonal frequency division multiplexing}
\acrodef{PDP}{power delay profile}
\acrodef{QAM}{quadrature amplitude modulation}
\acrodef{QPSK}{quaternary phase shift keying}
\acrodef{SINR}{signal to interference plus noise ratio}
\acrodef{SIR}{signal to interference ratio}
\acrodef{SNR}{signal to noise ratio}
\acrodef{SVD}{singular value decomposition}
\acrodef{ZS}{zero pad suffix}
\acrodef{iid}{independent and identically distributed}
\acrodef{pdf}{probability density function}

\newcommand{\tr}[1]{\mathrm{tr} \left( #1 \right)}

\newcommand{\toep}[2]{{\mathop{\rm Toep}\nolimits}_{#1}\left(#2\right)}

\begin{document}

\title{Low-power Secret-key Agreement over OFDM}
%
\author{\IEEEauthorblockN{Francesco Renna\IEEEauthorrefmark{1}, Nicola Laurenti\IEEEauthorrefmark{2}, Stefano Tomasin\IEEEauthorrefmark{2}, Marco Baldi\IEEEauthorrefmark{3} \\ Nicola Maturo\IEEEauthorrefmark{3}, Marco Bianchi\IEEEauthorrefmark{3}, Franco Chiaraluce\IEEEauthorrefmark{3}, and Matthieu Bloch\IEEEauthorrefmark{4}}\\
\IEEEauthorblockA{\IEEEauthorrefmark{1}%
Instituto de Telecomunica\c{c}\~{o}es, Departamento de Ci\^encia de Computadores, 
Universidade do Porto,\\ \texttt{frarenna@dcc.fc.up.pt}}\\[-2mm]
\IEEEauthorblockA{\IEEEauthorrefmark{2}%
Department of Information Engineering, University of Padua,\\ \texttt{\{nil, tomasin\}@dei.unipd.it}}\\[-2mm]
\IEEEauthorblockA{\IEEEauthorrefmark{3}%
Department of Information Engineering, Universit\`a Politecnica delle Marche, \\ \texttt{\{m.baldi, n.maturo, m.bianchi, f.chiaraluce\}@univpm.it}}\\[-2mm]
\IEEEauthorblockA{\IEEEauthorrefmark{4}%
School of Electrical Engineering, Georgia Institute of Technology, GT-CNRS UMI 2958\\ \texttt{matthieu.bloch@ece.gatech.edu}}}
%

\maketitle 

\begin{abstract}
Information-theoretic secret-key agreement is perhaps the most practically feasible mechanism that provides unconditional security at the physical layer to date. 
In this paper, we consider the problem of secret-key agreement by sharing randomness at low power over an \ac{OFDM} link, in the presence of an eavesdropper.
The low power assumption greatly simplifies the design of the randomness sharing scheme, even in a fading channel scenario. We assess the performance of the proposed system in terms of secrecy key rate and show that a practical approach to key sharing is obtained by using \ac{LDPC} codes for information reconciliation. Numerical results confirm the merits of the proposed approach as a feasible and practical solution. Moreover, the outage formulation allows to implement secret-key agreement even when only statistical knowledge of the eavesdropper channel is available.
\end{abstract}


\section{Introduction}
Wireless communication systems and networks are particularly prone to attacks, because the inherent broadcast nature of the radio channel makes any terminal in the transmission range a potential threat. Physical-layer security aims at strengthening these systems by exploiting the imperfections of communication channels with appropriate coding and signaling strategies at the physical layer. 
Since the seminal works \cite{Wyner,Maurer,AhlCsi}, physical-layer security has mainly focused on two mechanisms: secret communication over the wiretap channel, and secret-key agreement with the aid of a public side channel. 

While several results have established the benefits of diversity, fading \cite{Lai2012}, and multiple antenna \cite{Bloch2008,sktcom}, to improve secret-key rates over wireless channels, little has been done to analyze secret-key agreement in the context of \ac{OFDM} systems, which have 
become the reference wireless physical layer technique for high data rate wireless communications.
Previous work on secret-key agreement in \ac{OFDM} systems and fading environments has used a \emph{source model} for secret-key agreement based on channel reciprocity \cite{Liu2012}, so that separate measurements of the channel coefficients of the wireless link between the legitimate terminals could be used as the shared randomness. Other related works have also considered the problem of secure communication over \ac{OFDM} channels by modeling them as parallel wiretap channels \cite{Li2006,Jorswieck2008a}, based on the fact that an \ac{OFDM} system is designed to avoid interference among subchannels and among symbols. However, a sophisticated eavesdropper may refuse to implement the canonical \ac{OFDM} receiver, keeping the \ac{CP} samples to increase the amount of information he can get out of it, and thus creating interference among the wiretap channels. 

In this paper, we consider the problem of generating secret cryptographic keys over a wireless channel by using \ac{OFDM} transmission with \ac{CP} for randomness sharing. We consider a \emph{channel model} for secret-key agreement, in which randomness is injected into the channel by one of the legitimate terminals. Moreover, we analyze slow fading dispersive channels, for which the channel impulse responses are assumed to remain constant over the whole duration of the randomness sharing phase. We first show that in the low-power limit the strategy to allocate all transmit power on the subchannel having the highest channel gain to the legitimate receiver is first-order optimal. We derive the secret key achievable rates in this case, and observe that first-order optimality is retained by replacing Gaussian inputs with a \ac{QPSK} constellation with the same variance. Then, as a practical solution, we propose the use of \ac{LDPC} coding for information reconciliation. Indeed, \ac{LDPC} codes are state-of-the-art error correcting codes, characterized by soft decoding algorithms able to approach the unconstrained channel capacity, with limited complexity. They have already found several applications in physical-layer security, either as codes for near-optimal information reconciliation in secret-key agreement schemes \cite{Bloch2006, Elkouss2009, Wong2011}, or as codes for secure communication over wiretap channels \cite{Klinc2011, Baldi2010}. In order to assess the merit of the practical solution we also consider as performance metric the \textit{security gap} defined as the ratio between the legitimate and eavesdropper \ac{SNR} which allows reliable decoding for the legitimate receiver, while keeping the eavesdropper \ac{BER} and \ac{FER} sufficiently close to $0.5$ and $1$, respectively. We focus on regular \ac{LDPC} codes, since both their optimization and implementation are simpler than for irregular codes. In addition, regular \ac{LDPC} codes also include several classes of structured codes \cite{Baldi2011a}, which are well suited to practical implementation \cite{Baldi2009a}. Lastly, we discuss the design of the system based on an outage approach, when the channel of the eavesdropper is known only statistically to the legitimate transmitter.

We denote vectors and matrices with lowercase and uppercase boldface letters, respectively, and the complex conjugate transposed of matrix $\B A$ as with $\B A^\ast$. The eigenvalues of an $L\times L$ matrix $\B A$ are denoted by $\lambda_i(\B A), i = 1,\ldots,L$. Given a vector $\B g\in \M C^L$, we denote by $\toep N{\B g}$ the $(N+L-1)\times N$ Toeplitz matrix having $[g_1,0,\ldots,0]$ as its first row and $[g_1,\ldots,g_L,0,\ldots,0]$ as its first column.

\section{System model}

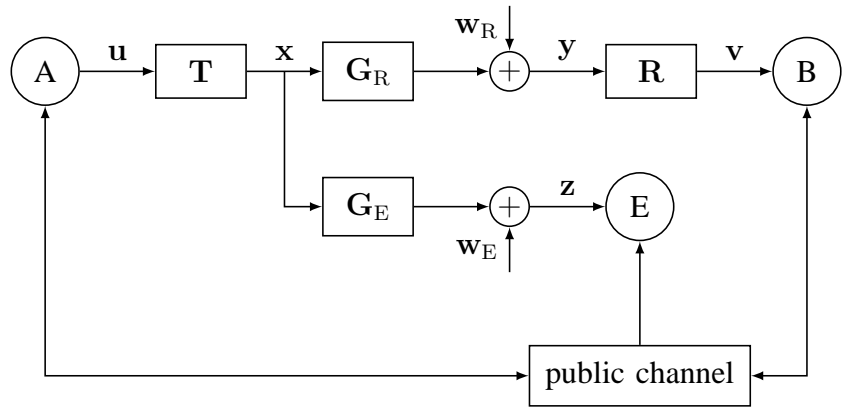
\begin{figure}
\begin{center}
\begin{tikzpicture}[x=1cm,y=1.5cm,semithick]
\draw (0,0) node(sou)[draw,circle,minimum size=9mm]{A};
\draw[-latex] (sou.east) --++(1,0) node[midway,above]{$\B u$}
node(mod)[draw,inner sep=6pt,minimum width=1.2cm,anchor=west]{$\B T$};
\draw[-latex] (mod.east) --++(1,0) node(x)[midway,above]{$\B x$}
node(hR)[draw,inner sep=6pt,minimum width=1.2cm,anchor=west]{$\B G\sub R$};
\draw[-latex] (hR.east) --++(1,0) 
node(sum)[draw,circle,inner sep=1pt,anchor=west]{$+$};
\draw[-latex] (sum.east) --++(1,0) node[midway,above]{$\B y$}
node(dem)[draw,inner sep=6pt,minimum width=1.2cm,anchor=west]{$\B R$};
\draw[-latex] (dem.east) --++(1,0) node[midway,above]{$\B v$}
node(use)[draw,circle,minimum size=9mm,anchor=west]{B};
\draw[latex-] (sum.north) --++(0,0.4) node[midway,left]{$\B w\sub R$};
\draw[-latex] (x.south) |-++(0.5,-1.2) 
node(hJ)[draw,inner sep=6pt,minimum width=1.2cm,anchor=west]{$\B G\sub E$};
\draw[-latex] (hJ.east) --++(1,0) 
node(sumj)[draw,circle,inner sep=1pt,anchor=west]{$+$};
\draw[-latex] (sumj.east) --++(1,0) node[midway,above]{$\B z$}
node(eve)[draw,circle,minimum size=9mm,anchor=west]{E};
\draw[latex-] (sumj.south) --++(0,-0.4) node[midway,left]{$\B w\sub E$};
\draw (eve) ++ (0,-1.5) node(pc)[draw,inner sep=6pt,minimum width=1.2cm]{public channel};
\draw [latex-latex] (sou) |- (pc);
\draw [latex-latex] (use) |- (pc);
\draw [latex-] (eve) -- (pc);
\end{tikzpicture}
\end{center}
\caption{Block diagram for the vector/matrix representation of a secret key agreement scheme based on OFDM.}
\label{fig:schemaBlocchi}
\end{figure}

In the typical physical-layer key-agreement scenario, two legitimate terminals, which we call A and B, aim at deriving a common bit sequence (the key) that must be kept secret from an adversary who will be called E. For this purpose, A and B have access to a noisy wireless link and to a public, error-free authenticated channel. However, it is assumed that, due to the nature of the wireless medium, a link also exists from A to E and that messages on the public channel may be observed by E.  

We consider that the wireless link is implemented through an \ac{OFDM} system with $M$ subcarriers, equally spaced in frequency, and a \ac{CP} of $\mu$ samples.
For convenience, we use the matrix representation of the \ac{OFDM}/CP system introduced in \cite{RennaTIFS12}, that can be inferred from Fig.~\ref{fig:schemaBlocchi}. The description is based on the discrete time equivalent of the system with $N$ samples per symbol period, and its efficient implementation through the \ac{FFT} algorithm. 
We assume that the \ac{CP} is longer than the main channel impulse response ${g}\sub R$ in order to avoid \ac{ISI} and \ac{ICI} at the legitimate receiver. 
The input-output relationships are then a special case of the \ac{MIMO} Gaussian wiretap channel:
\begin{equation}
\begin{array}{r@{\;=\;}l}
 \B{y}&  \B{G}\sub R \B{x}+\B{w}\sub R\\[1mm]
  \mbox{and\quad} \B{z}  &  \B{G}\sub E \B{x}+\B{w}\sub E 
\end{array}
 \label{eq:MIMOwiretap}
\end{equation}
where the vector $\B{x} \in \mathbb{C}^{N}$ contains the signal samples corresponding to an \ac{OFDM} symbol, transmitted on the  channel, while multiplications of $\B{x}$ by the Toeplitz matrices 
$\B{G}\sub R = \toep N{\B g\sub R}$ and $\B{G}\sub E = \toep N{\B g\sub E}$
are the convolutions of the input signal with the channel impulse responses $\B{g}\sub R = \left[g\sub R(0),\ldots,g\sub R(L\sub R-1)\right]$ and $\B{g}\sub E = \left[g\sub E(0),\ldots,g\sub E(L\sub E-1)\right]$, having lengths $L\sub R$ and $L \sub E$, respectively.
The noise vectors ${\B{w}\sub R,\B{w}\sub E} \sim  \mathcal{C\! N}(0,\B{I}_{N+L_i-1})$, with $i = \mathrm{R},\mathrm{E}$, {comprise }independent, zero-mean, {unit-variance, }circularly symmetric complex Gaussian {variables}. 

To impose the \ac{OFDM} structure on the transmitted signal, we write
\begin{equation}
\B{x}=\B{T} \B{u} ,
\label{eq:OFDMchannelinput}
\end{equation}
where the vector $\B{u} \in \mathbb{C}^{M}$ contains the frequency domain symbols loaded on the $M$ subcarriers. The \ac{OFDM} modulation matrix $\B{T}$ is an $N \times M$ matrix that can be written as $\B{T}=\B{A} \B{F}^*$, in which $\B{F}$ represents the \ac{FFT} matrix of size $M$, while $\B A\in\mathbb C^{N\times M}$ is responsible for inserting $\mu = N-M$ redundant samples that are needed to overcome the delay spread of the dispersive channel, i.e.,
\begin{equation}
\B{A}= 
\left[
\begin{array}{ccccccc}
  {\B{0}}   & \B{I}_{\mu} \\
\hline
\multicolumn{2}{c}{\B{I}_{M}}
\end{array}
\right].
\label{eq:Tcp}
\end{equation}
{Similarly, } demodulation at the receiver is represented by the multiplication $\B v = \B R \B y$ of the legitimate channel output by the matrix $\B{R}=\B{F} \B{B}$. Here $\B{B}$ is such that under the condition $L\sub R \leq \mu$, 
\begin{equation}
\B{R} \B{G}\sub R \B{T}=\mathrm{diag}(\mathcal{G}\sub R(f_i)), 
\end{equation}
in which $\mathcal{G}\sub R(f_i)$, $i=1,\ldots,M$ is the length $M$ \ac{FFT} of the legitimate channel impulse response.
Thus,
\begin{equation}
  \B{B}=
 \left[
 \begin{array}{lll}
 \B{0}_{M \times \mu} & \B{I}_M &  \B{0}_{M \times (L\sub R-1)} 
 \end{array}
 \right]\ .
\label{eq:Rcp}
\end{equation}
Given the above, the \ac{OFDM} system scenario with generic eavesdropper can be represented as an equivalent \ac{MIMO} Gaussian wiretap channel:
\begin{equation}
\begin{array}{r@{\;=\;}l}
 \B{v}&  \B{H}\sub R \B{u}+\B{w}'\sub R\\[1mm]
  \mbox{and\quad} \B{z}  &  \B{H}\sub E \B{u}+\B{w}\sub E 
\end{array}
 \label{eq:OFDM_MIMO}
\end{equation}
with ${\B{H}}\sub R=\mathrm{diag}(\mathcal{G}\sub R(f_i))$, ${\B{H}}\sub E=\B{G}\sub E \B{T}$ and $\B{w}'\sub R=\B{R}\B{w}\sub R$. Consequently, the covariance matrix of the demodulated noise at the legitimate receiver {is }$\B{K}_{\B{w}'\sub R}=\B{R}\B{R}^*= \B{I}_M$.

\section{Low power randomness sharing and achievable secret-key rates}
 
From known results regarding the \ac{MIMO} Gaussian wiretap model, the secret-key capacity with a given input covariance matrix $\mathbf{K}_{\mathbf{u}}$ is obtained with Gaussian inputs, and is given by \cite{sktcom}
\begin{IEEEeqnarray}{rCl}
 R & = & \log_2 \left|   \mathbf{I}  + \mathbf{K}_{\mathbf{u}}^{\frac{1}{2}}  \left(  \mathbf{H}\sub R^\ast \mathbf{H}\sub R + \mathbf{H}\sub E^\ast \mathbf{H}\sub E   \right) \mathbf{K}_{\mathbf{u}}^{\frac{1}{2}}  \right|  -  \log_2 \left|   \mathbf{I}  + \mathbf{K}_{\mathbf{u}}^{\frac{1}{2}}   \mathbf{H}\sub E^\ast \mathbf{H}\sub E    \mathbf{K}_{\mathbf{u}}^{\frac{1}{2}}  \right|.
\end{IEEEeqnarray}
On the other hand, from \cite[Proposition~2]{sktcom}, we also know that, in the low-power regime, i.e.\ when the available power $P$ goes to zero,  the optimal transmission strategy is to concentrate all the power along the eigenspace of the legitimate channel $\B H\sub R$ corresponding to the maximum eigenvalue, regardless of the eavesdropper's channel. In our case, the optimal input covariance matrix that satisfies the total power constraint
 \begin{equation}
\tr{\B{K_{x}}}= \mathrm{tr}(\B{T} \B{K_u} \B{T}^*) \leq P\ ,
\label{eq:traceConOFDM}
\end{equation}
 is diagonal, with only one nonzero entry, corresponding to the subcarrier that exhibits the maximum channel gain. Namely,
 \begin{equation}
 \mathbf{K}_{\mathbf{u}} = \frac{P}{1+ \rho} \mathbf{e}_m \mathbf{e}_m^\ast,
 \end{equation}
 in which $\rho = \mu/M$, $\{ \mathbf{e}_i\}$ is the canonical base of $\mathbb{R}^M$ and 
 \begin{equation}
 m = \arg \max_i |\lambda_i(\mathbf{H}\sub R^\ast\mathbf{H}\sub R)| = \arg \max_i |\mathcal{G}\sub R(f_i)|\;.
 \end{equation}
Accordingly, the secret-key rate achieved for $P > 0$ with the low-power optimal transmission strategy is
\begin{IEEEeqnarray}{rCl}
R & = & \log_2 \frac{   1 + \frac{P}{1+\rho} | \mathcal{G}\sub R(f_m)|^2  + \frac{P}{1+\rho}  \|  \mathbf{H}\sub E \mathbf{e}_m \|^2 }{1  + \frac{P}{1+\rho}  \|  \mathbf{H}\sub E \mathbf{e}_m \|^2} \nonumber \\
 & = & \log_2 \frac{1+\Lambda\sub R+\Lambda\sub E}{1+\Lambda\sub E} \label{eq:Rk}
\end{IEEEeqnarray}
where $\Lambda\sub R = \frac{P}{1+\rho} | \mathcal{G}\sub R(f_m)|^2$ and $\Lambda\sub E = \frac{P}{1+\rho}  \|  \mathbf{H}\sub E \mathbf{e}_m \|^2$ are the \ac{SNR} of the two channels, relative to the chosen subcarrier.

Moreover, by leveraging the low-power, first-order expansion of mutual information in~\cite{Palomar06}, the result in \cite[Proposition 2]{sktcom} can be extended to any complex input with the same covariance matrix, and independent real and imaginary components. Therefore, in the low-power regime, Gaussian signaling is no longer necessary to achieve the secret-key capacity of the channel, and A can transmit symbols from a discrete constellation (e.g., \ac{QPSK}) without incurring significant losses with respect to expression \eqref{eq:Rk} \footnote{An analogous result holds for low-power secrecy capacity of a MIMO Gaussian channel\cite{Gursoy2012}.}.

\section{Practical Solution}\label{sec:IR}

The choice of a \ac{QPSK} modulation for randomness sharing simplifies the design of the information reconciliation phase. Indeed, as reconciliation of continuous variables is not needed, it can be effectively implemented through standard soft decoding techniques of a binary code in an \ac{AWGN} channel with binary input. For instance, \cite{Wong2011} employs fixed \ac{LDPC} codes with syndrome transmission on the public feedback channel. 

In this section, as an alternative to the standard reconciliation scheme, we derive a suboptimal, still more convenient, practical approach. We first observe that since the transmitter chooses the best subchannel to the legitimate receiver, it is quite likely that $\Lambda\sub R > \Lambda\sub E$. 
Therefore, the proposed approach is to use the resulting wiretap channel (said to be stochastically degraded) to deliver a secret key created at the transmitter, without leveraging the presence of a public, noiseless, side channel for discussion. 

As a further step towards practice, we consider finite length codes. In this context, we aim for a looser notion of secrecy, based on the eavesdropper \ac{BER} rather than mutual information.   

We now focus on the use of \ac{LDPC} codes as secrecy codes for the wiretap channel. We observe that their behavior can be approximated by assuming that if the SNR working point $\Lambda$ is close to the decoding threshold $\Lambda\sub{th}$, a small decrease of $\Lambda$ would cause the code to be unable to correct the errors. On the other hand, a small increase of $\Lambda$ would allow to decode correctly all of them. The threshold $\Lambda\sub{th}$ will be derived in the following. Under the physical-layer security viewpoint, the ideal condition would be reached if an eavesdropper at $\Lambda\sub{E} = \Lambda\sub{th} - \epsilon$, with $\epsilon$ arbitrarily small, was unable to get any information on the received codewords, while the authorized receiver at $\Lambda\sub{R} = \Lambda\sub{th} + \epsilon$ can perfectly recover the message. In this case, the security gap $S_g = {\Lambda\sub{R}}/{\Lambda\sub{E}}$ needed to achieve the security and reliability conditions would be very small.

In order to approach the ideal condition, we can consider rather long codes together with scrambling \cite{Baldi2010,Baldi2011,Baldi2012}. Under the hypothesis that the scrambler can approach the perfect scrambling condition, as defined in \cite{Baldi2012}, the \ac{BER} is about half the \ac{FER}; so, the eavesdropper's performance is strongly affected by his degraded channel.

For the derivation of $\Lambda\sub{th}$ we consider the density evolution technique, whose core is represented by the following recursion \cite{Moon}:
\begin{equation}
\xi_i = \Psi^{-1}\left(\left(\Psi\left(\frac{2E_c}{\sigma^2} + (w\sub c - 1)\xi_{i-1}\right)\right)^{w\sub r - 1}\right)\,.
\label{eq:mu}
\end{equation}
In (\ref{eq:mu}), $\xi_i$ denotes the mean of a randomly chosen message from a check node in the associated Tanner graph at iteration $i$,
$E_c$ is the energy per codeword bit, $\sigma^2$ is the noise variance, $w\sub c$ and $w\sub r$ are the parity-check matrix column and row weights,
respectively, while the function $\Psi$ is defined as follows:
\begin{equation}
\Psi(x) = \frac{1}{\sqrt{4\pi x}} \int_{-\infty}^{+\infty} \mathrm{tanh}(y/2) e^{-\frac{(y-x)^2}{4x}} dy.
\end{equation}
The decoding algorithm is supposed to perform a maximum number of iterations equal to $I$.
If $\xi_i$ becomes greater than 1 for some $i \le I$, this means that the \ac{LDPC} code is able to correct all errors. Thus, by using \eqref{eq:mu}, we can obtain the maximum channel noise levels for which the message-passing decoder  will be able to correct all errors, which is also known as the decoding threshold for the specified ensemble  of \ac{LDPC} codes. For the ease of implementation, we use the approximated version of density evolution which assumes that all messages are Gaussian and also consistent (that is, with variance equal to twice the mean).

As an example, we have considered a QPSK modulated transmission over the \ac{AWGN} channel,
and an SNR working point $\Lambda = -2$\,dB.
Using density evolution, we find that $\Lambda\sub{th} \approx -2$\,dB for regular
\ac{LDPC} codes with: i) $w\sub c = 3$ and code rate $0.25$; ii) $w\sub c = 4$ and code rate $0.15$; iii) $w\sub c = 5$ and code rate $0.03$.
We have focused on $w\sub c = 3$, and used the progressive edge growth algorithm \cite{Hu2001} to design two (almost) regular \ac{LDPC} codes
with rate $0.25$ and length $5\,000$ and $10\,000$, respectively. With \ac{QPSK} modulation, the above code rate corresponds to 0.5 bits per channel use. 
Their performance has been assessed through numerical simulations, using the log-likelihood version of the 
sum-product algorithm \cite{Hagenauer1996} for \ac{LDPC} decoding.
\begin{figure}[tbh]
\begin{centering}
\includegraphics[keepaspectratio, width=0.7\columnwidth]{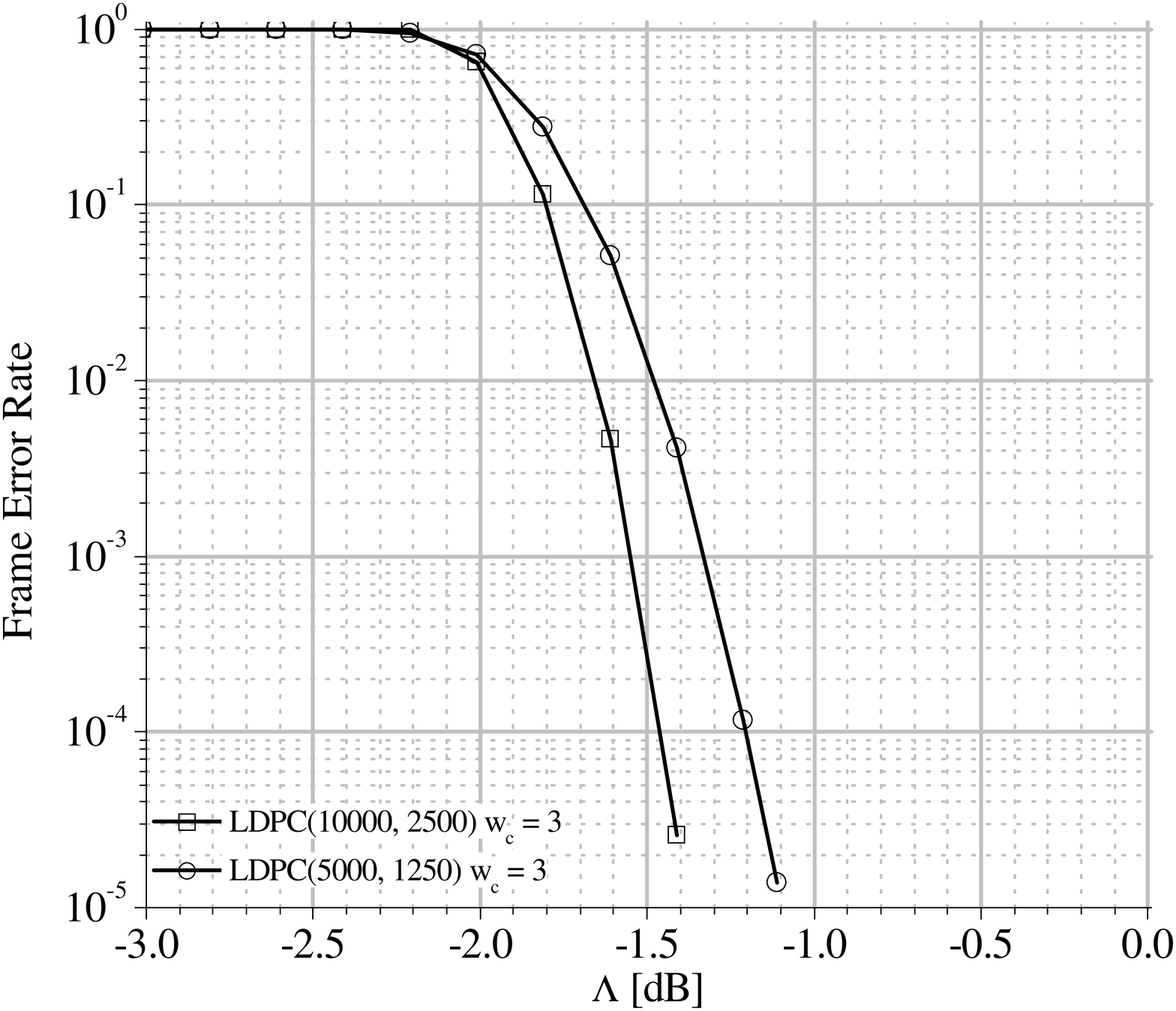}
\caption{Simulated frame error rate for two \ac{LDPC} codes with rate $0.25$, parity-check matrix column weight $3$, codeword length $5\,000$ and $10\,000$.}
\label{fig:LDPCsim}
\par\end{centering}
\end{figure}
The results obtained are reported in Fig. \ref{fig:LDPCsim}.
If we fix the security condition as to ensure that E experiences a \ac{FER} $\ge 0.9$, this is reached,
for both codes, when $\Lambda\sub{E} \le -2.2$\,dB.
Concerning B's reliability condition, we can require that the frame error rate he experiences is $\le 10^{-4}$.
This is achieved for $\Lambda\sub{R} \ge -1.2$\,dB, for the first code, and $\Lambda\sub{R} \ge -1.45$\,dB,
for the second code. Thus, the security gap is $S_g = 1$\,dB and $S_g = 0.75$\,dB, respectively.
Obviously, using longer codes would further reduce the security gap.

\section{Outage-based protocol design} \label{sec:OBPD}

While it seems reasonable to assume that the legitimate channel is perfectly known to both the legitimate terminals, and hence the optimal subcarrier index $m$ and the corresponding value $\Lambda\sub R$, assuming knowledge of the eavesdropper channel state is in general unrealistic. In the following, we assume that the transmitter only has statistical \ac{CSI} about the eavesdropper channel. 

\begin{figure} [tbh]
\begin{centering}
\includegraphics[keepaspectratio, width=0.9\columnwidth]{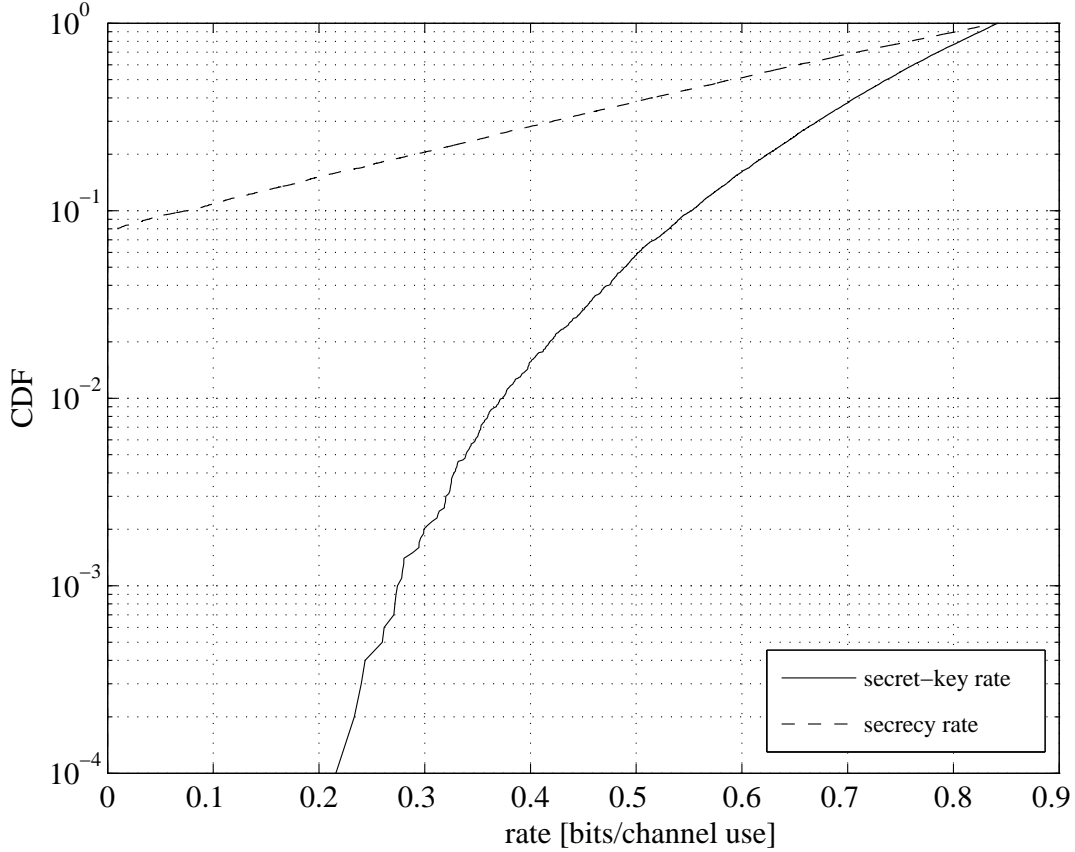}
\caption{CDF of the achievable secret key rates and secrecy rates when both the channels to B and E have an exponential \ac{PDP} with $\Gamma \sub R = \Gamma \sub E = -10\,{\rm dB}$, and the transmitted power $P$ is adjusted so that $\Lambda \sub R = -1\,{\rm dB}$.}
\label{fig:PGRminus1GEminus10}
\par\end{centering}
\end{figure}

\begin{figure} [tbh]
\begin{centering}
\includegraphics[keepaspectratio, width=0.9\columnwidth]{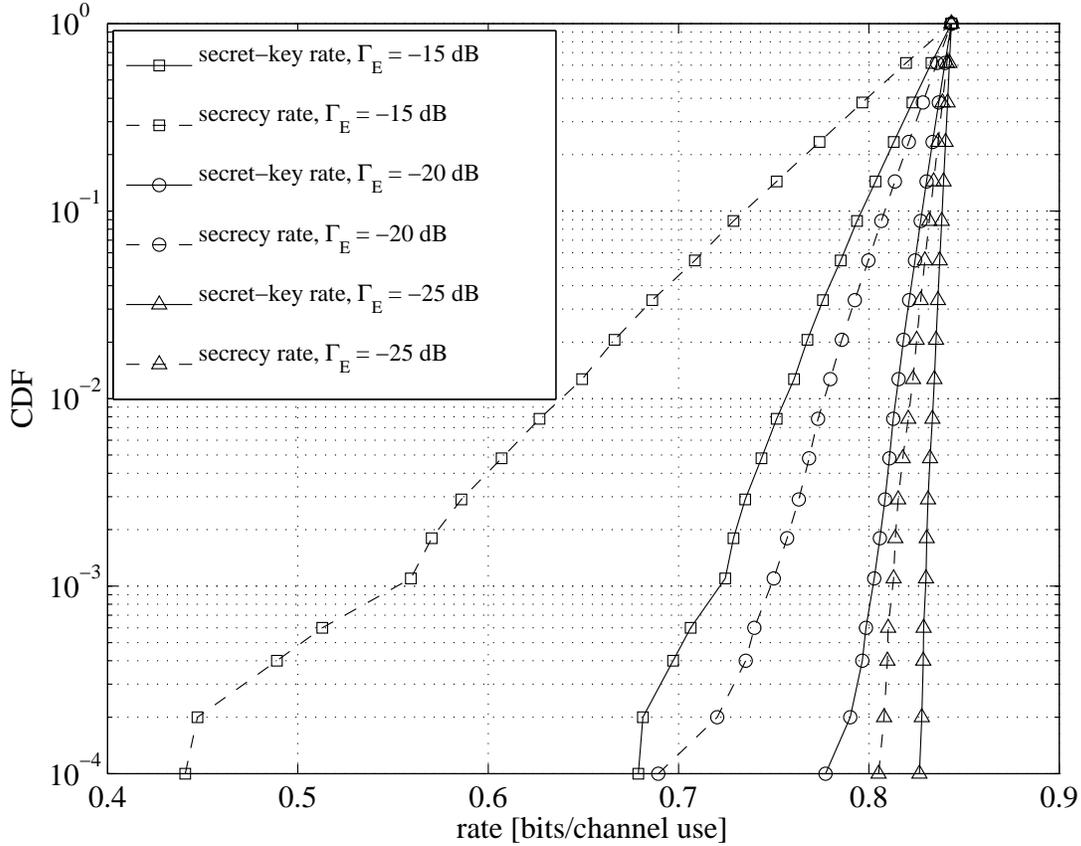}
\caption{CDF of the achievable secret key rates and secrecy rates when both the channels to B and E have an exponential \ac{PDP} with $\Gamma \sub R = - 10\,{\rm dB}$ and different values of $\Gamma \sub E$, and the transmitted power $P$ is adjusted so that $\Lambda \sub R = -1\,{\rm dB}$.}
\label{fig:GRminus1GEminus10}
\par\end{centering}
\end{figure}

The legitimate parties must therefore pursue a tradeoff between the key rate they settle for, and the secret key outage probability (that is the probability that the actual secret key capacity is lower than their intended rate). A possible approach is to always adjust the transmitted power $P$ so that $\Lambda\sub R$ has a fixed value. Then, the secret-key rate must be chosen so that the outage probability is small enough. 
An example is reported in Fig.~\ref{fig:PGRminus1GEminus10}, which illustrates the \ac{CDF} of the achievable secret key rates \eqref{eq:Rk} assuming the legitimate and eavesdropper channel coefficients are random realizations drawn from the same fading distribution. We considered an \ac{OFDM} system with $M=256$ subcarriers, \ac{CP} length $\mu = 16$, that is, transmitting over frequency selective channels with length $L \sub R = L \sub E = \mu$. Both the channel impulse responses towards B and E have independent Rayleigh fading taps with exponentially decaying \ac{PDP} and $\Gamma \sub R = \sum_i \mathbb{E}\left\{|g\sub R(i)|^2\right\} =  -10$\,dB and $\Gamma \sub E = \sum_i \mathbb{E}\left\{|g\sub E(i)|^2\right\} =  -10$\,dB. However, the transmission power $P$ is adjusted in order to guarantee $\Lambda\sub R = -1$\,dB. We see that by fixing an outage probability of $10^{-3}$ we should aim at a secret key rate $R = 0.28$ bit per channel use. For the sake of comparison, we also show in the figure the \ac{CDF} of the achievable secrecy rate for the same system and with the same input. Observe that the secrecy rate is always much lower than the secret-key rate with public discussion and may result in a zero rate with very high probability.
On the other hand, when the eavesdropper average \ac{SNR} is much lower than the one of the main channel, the achievable rates for the two schemes are quite close, as shown in 
Fig.~\ref{fig:GRminus1GEminus10}.

Notice that, in order to characterize the secret-key outage probability, it is important to determine the statistical description of the random variable $\Lambda\sub E$. In the Rayleigh fading case, as multiplying a complex Gaussian random variable by a constant phase term does not change its distribution, it can be seen that $\Lambda\sub E$ is distributed as
\begin{IEEEeqnarray}{rCl}
\tilde\Lambda\sub E &=& \frac{1}{{M}} \left[ \sum_{n = 1}^{L\sub E -1} \left(  \sum_{i=1}^n g\sub E(i) \right)^2 +  (N - L\sub E + 1 )  \left(  \sum_{i=1}^{L\sub E} g\sub E(i)  \right)^2  + \sum_{n = 1}^{L\sub E -1} \left(  \sum_{i=n+1}^{L\sub E} g\sub E(i) \right)^2 \right].
\end{IEEEeqnarray}
Then, $\tilde\Lambda\sub E$ can be easily rewritten as the quadratic form $\tilde\Lambda\sub E = \boldsymbol{\gamma}^\ast \mathbf{C} \boldsymbol{\gamma}$, in which $\boldsymbol{\gamma} \sim \mathcal{CN}(\mathbf{0},\mathbf{I}_{L\sub E})$ and $\mathbf{C}$ is a positive semidefinite matrix function of the system parameters $M, N$ and of the channel \ac{PDP}. Therefore, $\Lambda\sub E$ is distributed as the sum of independent exponential random variables with means equal to the eigenvalues of $\mathbf{C}$, $\lambda_1(\mathbf{C}), \ldots, \lambda_{L\sub E}(\mathbf{C})$. Then, the \ac{CDF} of $\Lambda\sub E$ is obtained as
\begin{equation}
F_{\Lambda\sub E}( \theta) =  \sum_{i=1}^{L\sub E} \frac{\lambda_i(\mathbf{C})^{L\sub E -1}}{\prod_{j \neq i} (\lambda_i(\mathbf{C})-\lambda_j(\mathbf{C})) }  \left( 1- e^{-\frac{\theta}{\lambda_i(\mathbf{C})}}  \right).
\end{equation}

Similarly, also when considering the practical solution based on the used of \ac{LDPC} codes for the wiretap channel scenario, described in Section \ref{sec:IR}, we can easily assess the effect of knowing E's channel only in statistical terms.
After having defined the security condition in terms of Eve's frame error probability, which implies $\Lambda\sub E < \Lambda\sub{th}-\varepsilon$, we can obtain the security outage probability as follows:
\begin{equation}
\mathbb{P} \left\{ \Lambda\sub E \ge \Lambda\sub{th}-\varepsilon \right\} = 1 - F_{\Lambda \sub E}(\Lambda\sub{th}-\varepsilon).
\label{eq:Po}
\end{equation}

\section{Conclusions}
We have considered the problem of information theoretical secret-key agreement by sharing randomness at low power over an \ac{OFDM} link, in the presence of an eavesdropper.

The low power assumption greatly simplifies the design and performance evaluation of the optimal scheme, even in a fading channel scenario and when the potential eavesdropper cannot be modeled as an \ac{OFDM} receiver.
In fact, by leveraging the analogy with a Gaussian MIMO channel, we have shown that the randomness sharing phase can be designed with complete ignorance of the eavesdropper channel state, without loss of optimality. It results in a QPSK modulation over the subcarrier that exhibits the maximum amplitude of the legitimate channel frequency response. 
As a further consequence, \ac{LDPC} codes, and their efficient soft decoding techniques can be employed effectively for information reconciliation, or directly as codes for the wiretap channel.

We have also provided an outage formulation and have explored the tradeoff between the secret key rate and the probability of secrecy outage, for proper dimensioning of the scheme.   

We point out that a similar approach can also be used for higher transmitted powers, although, in that case, a water-filling power distribution on the legitimate channel frequency response is suboptimal for randomness sharing. However, it is shown in \cite[Section V]{sktcom} to provide satisfactory results, and to achieve secret-key capacity again in the high power limit. The design of a protocol exploiting this solution will be pursued in our future work.

\section*{Acknowledgments}
This work was partially supported by the Italian Ministry for Education and Research (MIUR) project ``ESCAPADE'' (grant no. RBFR105NLC) under the``FIRB - Futuro in Ricerca 2010'' funding program.

\bibliographystyle{IEEEtran}
\bibliography{bibliography}

\end{document}